\documentclass[12pt]{article}
\usepackage{amsmath,amssymb,amsthm,amssymb}

\title{Violation of Bell's inequality and postulate on simultaneous measurement
of compatible observables}

\author{Andrei Khrennikov\\
International Center for Mathematical
Modelling \\in Physics and Cognitive Sciences \\ Linnaeus University, S-35195, V\"axj\"o, Sweden\\
Institute of Information Security\\
Russian State University for Humanities, Moscow, Russia}

\begin{document}

\maketitle

\abstract{We discuss coupling of violation of Bell's inequality
and \\ non-Kolmogorovness of statistical data in the EPR-Bohm
experiment. We emphasize that nonlocalty and ``death of realism''
are only sufficient, but not necessary
 conditions of non-Kolmogorovness. There
can be found other sufficient conditions of non-Kolmogorovness
and, hence,  violation of Bell's inequality. We find one important
source of non-Kolmogorovness by analyzing axiomatics of quantum
mechanics. We pay attention to the postulate (due to von Neumann
and Dirac) on simultaneous  measurement of quantum observables
given by commuting operators. This postulate is criticized as
nonphysical. We propose a new interpretation of the Born-von
Neumann-Dirac rule for calculation of the joint probability
distribution for such observables. It gets a natural physical
interpretation by considering conditional measurement scheme. We
use this argument (i.e., rejection of the postulate on
simultaneous measurement to motivate non-Kolmogorovness of the
probabilistic structure of the EPR-Bohm experiment.}

\section{Introduction}

\subsection{Nonlocality, ``death of reality''}
In the physical community violation of Bell's inequality \cite{B}
is typically considered as an evidence of either nonlocality or
``death of reality": local realism is incompatible with
predictions of QM. On the other hand, in the quantum logic and
quantum probability communities the same problem is often
interpreted in a completely different way. As was pointed out in
numerous papers (see, e.g., author's mongraphs \cite{KHR},
\cite{KHR1} and reviews \cite{KHR2}, \cite{KHR3} and see paper
\cite{Hess} in this issue),
 Bell's inequality can be violated simply because it is impossible
 to define a single probability measure which would serve for a few different
(incompatible)  experimental settings. The existence of such a
probability measure was  a hidden assumption in Bell's derivation
of his famous inequality. He assumed that all correlations can be
represented as averages with respect to the same probability
measure ($d\rho (\lambda)$ in his notations).

From the point of view of quantum logic/probability the EPR-Bohm
experiment does not have the  Kolmogorov probabilistic structure.
Hence, it could not be described by a single Kolmogorov
probability space.  If one works in the non-Kolmogorovian
framework, i.e., under the assumption that different experimental
settings induce their own probability measures, i.e., instead of a
single probability $d\rho (\lambda),$ one should consider a family
of probabilities $d\rho_{a, b} (\lambda)$ for settings $a, b,$
then only generalizations of Bell's inequality hold true
\cite{KHR}. Such generalized Bell's inequalities are not violated
by experimental data. However, the idea of non-Kolmogorovness of
the EPR-Bohm experiment did not propagate so much in the physical
community.

\medskip

Why was the message from quantum logic/probability  ignored in quantum physics?

\medskip

I think that one of the reasons  is that there was not presented a
natural physical mechanism of generating non-Kolmogorovness. One
should explain why Bell's identification \cite{B} of the
probability $d\rho (\lambda)$ with the probability distribution of
hidden parameters for the initial state can be questioned. One of
aims of this paper is to describe the process of generating
non-Kolmogorovness in the EPR-Bohm experiment.

\subsection{Simultaneous or conditional measurements}

The crucial point is that the  EPR-measurements should be
considered not as {\it simultaneous measurements on a pair of
entangled particles,} but as conditional measurements: first we
measure the projection of the polarization vector of the first
particle on the axis $a$ and then the projection of the
polarization vector of the second particle on the axis $b$ or vice
versa. Such a representation of the EPR-Bohm experiment was given
in the talk of Aspect at the Conference Foundations of Probability
and Physics-3 (V\"axj\"o University, June 7-12, 2004), see also
\cite{AA}. Unfortunately, he considered this shift from the
simultaneous measurement picture to the conditional measurement
picture just as a metaphor which is convenient for the
presentation of Bell's argument \cite{B}. Moreover, he used the
Copenhagen interpretation of the state vector: the $\psi$-function
gives the state of an {\it individual} quantum system (in the
EPR-Bohm case photon). Such interpretation immediately induces the
picture of nonlocal change of the state of e.g. the second
particle under the condition that the state of the first one was
collapsed in the process of measurement. In fact, in this
situation one gets nonlocality (under the assumption of realism)
automatically, without to appeal to Bell's argument. Alain Aspect
mentioned this in his talk; he considered Bell's inequality merely
as a test of the Copenhagen interpretation.

We shall use the so called {\it ensemble interpretation} of the
wave function:  the $\psi$-function describes statistical
properties of {\it an ensemble} of identically prepared quantum
systems. Therefore our conclusion will be completely different
from Aspect's conclusion.

We remind that Einstein had never accepted Bohr's thesis on
completeness of QM \cite{AE1}, \cite{AE2}.  All his life he
dreamed of creation of a new fundamental theory of micro
phenomena. He was sure that the wave function does not provide the
complete description of the state of an individual quantum system.
Einstein was the father of the {\it ensemble interpretation} of
the wave function as describing statistical properties of an
ensemble of systems created by some preparation procedure. This
interpretation was later elaborated by Leslie Ballentine
\cite{BL1}-\cite{BL3} who used the terminology the {\it
statistical interpretation.} Unfortunately, this terminology is
rather misleading, since it was used even by von Neumann: the wave
function, although assigned as the state of an individual system,
expresses statistics of measurements (but this statistics is
coupled to irreducible randomness).

We also remark that the construction in section 2 is similar to so
called filtering type measurements which were presented in very
detail in Ballentine's book \cite{BL2}. (Unfortunately, he did not
use our argument in Chapter 20 of his book: he presented only the
standard Bell argument.)

\section{Simultaneous measurement as an idealization of conditional measurements}

In this section we discuss a misconception that has been
propagating in the physics literature since the work of von
Neumann \cite{VN}. As is well known from linear algebra, two
diagonalizable matrices $a$ and $b$ can be diagonalized
simultaneously if and only if $a$ and $b$ commute. This theorem
says, of course, absolutely nothing about measurements.

This misconception is the source of a lot of confusion, wrong
statements, and ``paradoxes'' in the physics literature.  In
particular, it plays an important role in Bell's argument
\cite{B}.

One of the postulates of QM \cite{VN} tells  that if two
observables,  say $a$ and $b,$ are represented by commutative
operators, $\hat a$ and $\hat b: [\hat a, \hat b] = 0,$ then these
observables can be {\it measured simultaneously} (and vice versa).
The joint probability distribution is given by the Born (Dirac-von
Neumann) formula:
\begin{equation}
\label{*1}
P (a = \alpha, b = \beta) = ||P_\alpha^a P_\beta ^b \psi ||^2,
\end{equation}
where $P_\alpha^a$ and $P_\beta ^b$ are spectral projectors of
operators  $\hat a$ and $\hat b$ corresponding to eigenvalues
$\alpha$ and $\beta.$ (We restrict considerations to operators
with discrete spectra). This postulate has never been questioned
and it is commonly accepted (in contrast to e.g. von Neumann's
projection postulate).

I have doubts in validity of this postulate.  It seems that in
formula (\ref{*1}) there was encoded (by von Neumann and Dirac)
the probability distribution for the procedure of conditional
measurement. Von Neumann did not present a solid  physical
motivation of this postulate. He presented a rather long
consideration on a possibility to represent two self-adjoint and
commutative operators as functions of one fixed operator,
$$
= f(\hat{d}),\; \hat{b}= g(\hat{d}).
$$
Then he stressed that a measurement of the observable $d$
 provides automatically
measurements of observables $a$ and $b$  represented by operators
$\hat{a}$ and $\hat{b}:$ the value $d=\gamma$ is transferred into
the corresponding values $a=f(\gamma)$ and $b= g(\gamma).$

The ``hidden postulate'' of von Neumann which made his
construction quite natural from the physical viewpoint is that any
self-adjoint operator corresponds to some quantum observable -- in
the above consideration it was the correspondence $\hat{d} \mapsto
d.$  However, this hidden postulate by itself is not so natural
from the physical viewpoint. We also remark that this postulate
plays an important role in von Neumann's no-go theorem -- the
first no-go theorem \cite{VN}.

The main physical reason to reject the postulate  on the
possibility of simultaneous measurement is impossibility to
realize such a measurement in the real experimental setup - at
least for measurements on composite systems.

Let us consider the EPR-Bohm experiment for measurement of
projections onto axes $a$ and $b$ of polarizations for pairs of
entangled photons. It is well known that in real experiments $a$
and $b$ are not measured simultaneously. There is so called time
window $\Delta$, see, e.g., \cite{Weihs}, \cite{KHR1}, which plays
the fundamental role in forming of the probabilistic data; cf.
\cite{Raedt}--\cite{Raedt2}; see also \cite{KHR6} (coupling of
time-synchronization in the EPR-Bohm experiment with the use of
the projection postulate). The crucial point is that in this
experiment $\Delta$ could not be chosen arbitrary small! First of
all, if one were so naive to put $\Delta = 0,$ then there would be
no matched clicks of detectors at all. But even if $\Delta >0,$
but it is small then the majority of entangled pairs (which are
generated by a source) disappear. Thus the picture of Alain Aspect
is correct: first measurement on one photon and with some
nontrivial delay on the second. This is not at all the
simultaneous measurement which was discussed by Dirac and von
Neumann, see, e.g., \cite{VN}.

Thus one might reject the very idea of simultaneous measurement as
nonphysical. In such a case one should provide a reinterpretation
of the formula (\ref {*1}) in the conditional probabilistic
framework. It can be easily done by using the framework of quantum
conditional probability \cite{KHR1}.

Let us consider two observables $a$ and $b$ which  are represented
by self-adjoint operators $\hat a$ and $\hat b$ with purely
discrete spectrum. At the moment commutativity is not assumed.
There is the initial state, say $\psi.$ The a was measured and the
result $a = \alpha$ occurred. Then the initial state $\psi$ is
transferred into the post measurement state
\begin{equation}
\label{*2}
\psi_\alpha = \frac{P_\alpha^a \psi}{||P_\alpha^a \psi ||.}
\end{equation}

\medskip

{\bf Remark.} We remark that, although this transformation is
commonly used in QM, especially, in quantum information theory,
its applicability can be questioned, see \cite{KHR4}, \cite{KHR5}
(especially, for measurements on composite systems; in particular,
in the EPR-type experiments). As was pointed out in \cite{KHR4},
\cite{KHR5}, in the case of observables with degenerate spectra
von Neumann did not define the post-measurement state  by
(\ref{*2}), see \cite{VN}. By von Neumann even for a pure initial
state $\psi$ the post-measurement state need not be again a pure
state again, it can become a mixture, i.e., it can be given not by
a state vector, but by a density matrix. Nowadays this von
Neumann's viewpoint is practically forgotten. In \cite{KHR4},
\cite{KHR5} was shown that incompatibility of von Neumann's
projection postulate and (\ref{*2}) for observables with
degenerate spectra induces an important objection to the standard
treatement of the EPR-experiment. We remark that (\ref{*2}) is
given by the L\"uders projection postulate. In the present paper
we do not discuss this delicate point; we proceed as it is usually
done in QM (recently the author demonstrated \cite{KHR7} that
under sufficiently general experimental conditions one can really
proceed with the L\"uders projection postulate; it seems that the
EPR-Bohm experiment satisfies conditions of \cite{KHR7}).

\medskip

We now would like to measure the $b$-observable. The crucial point
is that it is measured not for the initial state $\psi$ (not for
the initially prepared ensemble $S_\psi$), but for the post
measurement state $\psi_\alpha$ (for the  ensemble of those
systems for which the result $a = \alpha$ was obtained). The
conditional probability
\begin{equation}
\label{*3}
P_\psi (b = \beta | a = \alpha) \equiv P_{\psi_\alpha} (b= \beta) = \frac{||P_\beta^b P_\alpha^a \psi ||^2}{||P_\alpha^a \psi ||^2}.
\end{equation}
We now recall that
\begin{equation}
\label{*4}
|| P_\alpha^a \psi ||^2 = P_\psi (\alpha = a)
\end{equation}
(the probability to get the value $a = \alpha$ for a system belonging to the initial ensemble $S_\psi$).
Thus
$$|| P_\beta^b P_\alpha^a \psi ||^2 = P_\psi (a = \alpha) P_\psi (b = \beta \vert a = \alpha) = Q_\psi (a = \alpha, b = \beta).$$
The latter probability is the probability to get first the value
$a = \alpha$ and the then value $b = \beta.$  Thus one may
interpret Born-Dirac-von Neumann formula ({\ref{*1}}) as the rule
to find joint probability not for simultaneous measurement, but
for sequential measurement of $a$ and then $b$.

Let us now repeat previous consideration by changing the order of
measurements of $a$ and $b$. First we measure $b$. The probability
to get the value $b = \beta$ is given by $P_\psi (b = \beta) =
||P_\beta^b \psi ||^2$. The occurrence of this result induces a
new quantum state:
$$\psi_\beta = \frac{P_\beta^b \psi}{|| P_\beta^b \psi||}$$
and a new ensemble $S_{\psi_\beta}$ of  systems is created via
filtration with respect to this value. We can now perform the
$a$-measurement for systems belonging $S_{\psi_\beta}$ and find
the conditional probability $$P_\psi (a= \alpha \vert b =\beta)
\equiv P_{\psi_\beta} (a = \alpha) = \frac{|| P_\alpha^a P_\beta^b
\psi ||^2}{||P_\beta^b \psi ||^2}.$$ Thus
$$|| P_\alpha^a P_\beta^b \psi ||^2 = P_\psi (b = \beta) P_\psi (a = \alpha \vert  b = \beta) = Q_\psi (b = \beta, a = \alpha),$$
where by our conditional interpretation the  latter probability is
the joint probability of the sequential measurement: first
$b=\beta$ and then $a = \alpha.$ If
$$
Q_\psi(a = \alpha, b = \beta) = Q_\psi ( b=\beta, a = \alpha),
$$
one could forget about the order of measurements. Commutativity of
operators $\hat a$ and $\hat b$ is a {\it sufficient condition} of
such a coincidence, (\ref{*2}). It seems that it was the main
reason for invention of the Dirac-von  Neumann postulate on
simultaneous measurement of observables which are represented by
commutative operators. Commutative induced impression that, since
one need not take care of the order of measurements, it is
possible to interpret sequential probabilities $Q_\psi (a =
\alpha, b = \beta)$ and $Q_\psi (b = \beta, a = \alpha)$ as just a
single probability
\begin{equation}
\label{*3a} P_\psi (a = \alpha, b = \beta) \equiv Q_\psi (a =
\alpha, b = \beta) = Q_\psi (b = \beta, a = \alpha),
\end{equation}
i.e., that there exists a probability measure $P_\psi$ which does
not depend on measured observables $a$ and $b$ represented by
commutative operators and such that all bi-measurement
probabilities can be represented on the basis of this single
measure.

\section{EPR-Bohm experiment}

If we take another pair of observables, say $c$ and $d,$
represented by commutative operators  $\hat c$ and $\hat d,$ we
might misleadingly operate with the probability of simultaneous
measurement, $P_\psi ( c =\gamma, d=\epsilon).$

This induces the impression that all  such probability
distributions are related to the same Kolmogorov probability
measure $P_\psi.$ And it would be correct if one uses the
simultaneous measurement interpretation of probabilities under
consideration. However we use the conditional measurement
interpretation. Here the probability $Q_\psi (a = \alpha, b =
\beta)$ is based not only on  probability with respect to the
original state $\psi,$ namely, $P_\psi (a = \alpha),$ but also on
probability with respect to a completely different state, namely,
$P_{\psi_\alpha} (b=\beta).$

In the same way $Q_\psi (c = \gamma, d = \epsilon)$ is based not
only on  the $\psi$-probability $P_\psi (c = \gamma),$ but also on
the $\psi_\gamma$-probability $P_{\psi_\gamma} (d = \epsilon).$

\medskip

{\it Probabilities $P_{\psi_\alpha}$ and $P_{\psi_\gamma}$ need
not coincide.}  Therefore $Q_\psi (a = \alpha, b = \beta)$ and
$Q_\psi (c=\gamma, d=\epsilon)$ could not be represented as
probability distributions with respect to a single probability
measure.

\section{The conditional probabilistic structure of the EPR-Bohm experiment}
In the derivation of Bell's inequality \cite{B}
$$|\langle a^{(1)}, b^{(2)} \rangle - \langle b^{(1)}, c^{(2)} \rangle| \leq 1 - \langle a^{(1)}, c^{(2)}\rangle$$ Bell used a single probability measure $d\rho(\lambda).$ Here the indexes 1 and 2 are related to measurements on the first and the second particle, respectively, in the EPR pair of photons; $a, b, c$ are orientations of polarization beam splitters. By Bell's ``hidden assumption'' \cite{B}:
 $$
\langle a^{(1)}, b^{(2)} \rangle = \int_\Lambda a^{(1)} (\lambda) b^{(2)} (\lambda) d\rho (\lambda), \ldots,
\langle a^{(1)}, c^{(2)} \rangle = \int_\Lambda a^{(1)} (\lambda) c^{(2)} (\lambda) d\rho (\lambda).$$

To compare classical correlations with quantum mechanical
correlations, J. Bell assumed the validity of the Dirac-von
Neumann postulate on the simultaneous measurement. We can say that
this postulate was ``super-hidden assumption''. People never paid
attention on its crucial role in Bell's argument.

Now we consider the same EPR-Bohm experiment not from the
viewpoint of  simultaneous measurements of projections of
polarizations (which is evidently nonphysical), but from the
view-point of conditional (sequential) measurements (which
corresponds to the real experimental situation).

Consider the $(a^{(1)}, b^{(2)})$-measurement. The results of
measurements can be divided into three groups: $G_{12} (a^{(1)},
b^{(2)})$ first $a$ clicks for the first particle and only then
$b$ clicks for the second one, $G_{21} (a^{(1)}, b^{(2)})$ - vice
versa, $G(a^{(1)}, b^{(2)})$ - simultaneous clicks.

Since the number of simultaneous clicks is negligible we can
forget about $G(a^{(1)}, b^{(2)})$ and operate with only
$G_{12}(a^{(1)}, b^{(2)})$ and $G_{21}(a^{(1)}, b^{(2)}).$ The
group $G_{12}(a^{(1)}, b^{(2)})$ can be split into two subgroups
$G_{12}^\alpha (a^{(1)}, b^{(2)}), \alpha = \pm 1,$ corresponding
to results of measurements on the first photon: $a^{(1)} =
\alpha.$ We should associate with each such subgroup its own
probability measure $d P_\alpha^{a^{(1)}} (\lambda), \alpha = \pm
1.$

We point out that the probability $P_\alpha^{a^{(1)}}$ does not depend on $b^{(2)}.$
 This is the {\it condition of locality}. Quantum mechanics is
 local in the conditional measurement framework.

 Finally, as J. Bell did, we also consider the initial distribution of
 hidden variables $dP^0 (\lambda)$ corresponding the initial state preparation. The crucial point is
 that the covariance $\langle a^{(1)}, b^{(2)} \rangle$ could not be expressed in terms of only $dP^0 (\lambda),$
 the probabilities $dP_\alpha^{a^{(1)}}(\lambda)$ should be involved:
$$\langle a^{(1)}, b^{(2)} \rangle$$
$$
= P^0 (\lambda \in \Lambda: a^{(1)} (\lambda) = +1) \int_\Lambda b^{(2)} (\lambda) dP_+^{a^{(1)}} (\lambda)
$$
$$
- P^0 (\lambda \in \Lambda: a^{(1)} (\lambda) = -1) \int_\Lambda b^{(2)} (\lambda) dP_-^{a^{(1)}} (\lambda).
$$
If we repeat the previous considerations for the pair of settings ($b^{(1)}, c^{(2)}$) we obtain:
$$
\langle b^{(1)}, c^{(2)} \rangle
$$
$$
= P^0 (\lambda \in \Lambda: b^{(1)}(\lambda) = + 1) \int_\Lambda c^{(2)} (\lambda) dP_+^{b^{(2)}} (\lambda)
$$
$$
- P^0 (\lambda \in \Lambda: b^{(1)} (\lambda) = -1) \times \int_\Lambda c^{(2)} (\lambda) dP_-^{b^{(1)}} (\lambda).
$$
Probabilities $P_\pm^{a^{(1)}}$ and $P_\pm^{b^{(1)}}$ can differ.
Therefore one is not able to repeat  manipulations which had been
done by J. Bell. There is no Bell's inequality and, hence, no
problems at all.\footnote{By operating with a family of
probabilities one can derive generalized Bell's inequalities
\cite{KHR}. But such inequalities do not contradict to
probabilistic predictions of QM.}

\section{Discussion}

We demonstrated that Bell's arguments  were fundamentally based on
the Dirac-von Neumann postulate on the possibility of {\it
simultaneous measurement} of observables represented by
commutative operators. This QM-postulate was projected onto a
prequantum model with hidden variables. Consequently Bell assumed
that (classical) observables in all pairs:
$$
(a^{(1)}(\lambda), b^{(2)}(\lambda)), \; (b^{(1)} (\lambda), c^{(2)} (\lambda)), \; (d^{(1)} (\lambda), c^{(2)} (\lambda)),
$$
can be measured simultaneously. This assumption induces the illusion that covariations for all these pairs of classical observables, namely, $$
\langle a^{(1)}, b^{(2)} \rangle,  \; \langle b^{(1)}, c^{(2)} \rangle,\;  \langle a^{(1)}, c^{(2)} \rangle
$$
can be written with respect to a single probability measure: $d
\rho (\lambda)$ in Bell's  notations ($d P^0 (\lambda)$ in our
notations) corresponding the initial state $\psi.$ This assumption
induces the derivation of Bell's inequality. Violation of the
latter implies the revolutionary conclusion that QM is
incompatible with local realism.

We propose to reject the Dirac-von Neumann postulate on
simultaneous measurement. We propose to  interpret the
Born-Dirac-von Neumann formula (\ref{*1}) for the probability
distribution for simultaneous measurement as the formula for the
joint probability distribution in the sequential measurement. This
formula can be applied even in the case of observables represented
by noncommutative operators. The main motivation of our
substitution of the postulate on conditional measurements in the
place of the postulate on simultaneous measurement is the evident
experimental fact that simultaneous measurement is really
impossible in experiments with composite systems, e.g., pairs of
entangled photons. The time window is always nontrivial.
Measurements are always conditional (sequential). Therefore
conditional probabilistic formalism should be applied. Finally, we
remark that the rejection of the Dirac-von Neumann postulate on
simultaneous measurement induces just minority reconsideration of
foundations of physics comparing with rejection of local realism
(or as an alternative - questioning of the validity of the
mathematical formalism of QM). \footnote{We remark that the
Dirac-von Neumann postulate does not belong to the mathematical
domain of QM. It belongs to the interpretation of the mathematical
formalism of QM. We proposed another interpretation for the
mathematical formula (\ref{*1}).}

We recall that our representation of the  EPR-Bohm experiment as
conditional measurement recalls the original consideration of EPR
\cite{EPR}. There was nothing about simultaneous measurement in
the original EPR-framework. J. Bell by introducing simultaneous
measurements changed the original problem. He analyzed the
EPR-Bohm experiment under the additional assumption of validity of
the Dirac-von Neumann postulate. He did not recognized the
fundamental role of this assumption in his model of prequantum
reality. Of course, the realization of ``Bell's project" has a
fundamental consequence that the Dirac von Neumann postulate
should be rejected. However, Bell did not recognize this and he
used this argument to support the hypothesis on nonlocality. (Bell
was "nonlocal realist".)

\medskip

I would like to thank W.M. de Muynck for fruitful discussion on
the conditional probability viewpoint on Bell's inequality and the
EPR-Bohm experiment as well as the role of the projection
postulate (and its two versions -- by von Neumann and by
L\"uders), see also \cite{Mu}; I also would like to thank L.
Ballentine for numerous discussions on the statistical
interpretation of QM.

This paper was completed during my visit to Tokyo University of Science, Kyoto University and Nagoya University, March-April  2010.
I would like to thank N. Watanabe, M. Ohya, I. Ojima and M. Ozawa for hospitality and interesting discussions.
My studies on quantum foundations were supported by the grant  (2006-2010) ``Mathematical Modeling and System Collaboration" of
Linnaeus University, V\"axj\"o-Kalmar, Sweden.

    1. C. D. Scott and R. E. Smalley, J. Nanosci. Naotech. 3, 75 (2003)
2. J. M. Cowley, Diffraction Physics, Elsevier, Amsterdam (1995)

3. W. Heimbrodt and P. J. Klar, in Magnetic Nanostructures, Edited
by H. S. Nalwa, American Scientific Publishers, Los Angeles
(2002), Chapter 1, pp.1-58

4. J. P. Turner and P. C. Chu, in Microelectronics, T. J. Kern,
Ed., Materials Research Society, Warrendale, PA (1995), Vol. 143,
p.375.

\end{document}